%Paper: gr-qc/9512049
%From: hu@umdhep.umd.edu
%Date: Fri, 29 Dec 1995 09:15:17 EST

%%                              JNL.TEX
%%
%%                This is JNL.TEX Version 0.3 as of 6/12/85.
%%
%%      This is a set of TeX 82 macros designed to produce scientific
%%      papers with a minimum of fuss and using as much of plain.tex as
%%      possible.  The user need only know what is in the TeXbook, and
%%      the macros under ``user definitions'' below.  Also, the user
%%      definitions are intended to be as simple as possible, so that
%%      the user may change them as desired.

%%
%%  Font definitions suitable for the IMAGEN (Written by Tony Kennedy)
%%

%  Define a whole menagerie of pseudo-12pt fonts

\font\twelverm=cmr10 scaled 1200    \font\twelvei=cmmi10 scaled 1200
\font\twelvesy=cmsy10 scaled 1200   \font\twelveex=cmex10 scaled 1200
\font\twelvebf=cmbx10 scaled 1200   \font\twelvesl=cmsl10 scaled 1200
\font\twelvett=cmtt10 scaled 1200   \font\twelveit=cmti10 scaled 1200

\skewchar\twelvei='177   \skewchar\twelvesy='60

%  Define \...point macros to change fonts and spacings consistently

\def\twelvepoint{\normalbaselineskip=12.4pt
  \abovedisplayskip 12.4pt plus 3pt minus 9pt
  \belowdisplayskip 12.4pt plus 3pt minus 9pt
  \abovedisplayshortskip 0pt plus 3pt
  \belowdisplayshortskip 7.2pt plus 3pt minus 4pt
  \smallskipamount=3.6pt plus1.2pt minus1.2pt
  \medskipamount=7.2pt plus2.4pt minus2.4pt
  \bigskipamount=14.4pt plus4.8pt minus4.8pt
  \def\rm{\fam0\twelverm}          \def\it{\fam\itfam\twelveit}%
  \def\sl{\fam\slfam\twelvesl}     \def\bf{\fam\bffam\twelvebf}%
  \def\mit{\fam 1}                 \def\cal{\fam 2}%
  \def\tt{\twelvett}
  \textfont0=\twelverm   \scriptfont0=\tenrm   \scriptscriptfont0=\sevenrm
  \textfont1=\twelvei    \scriptfont1=\teni    \scriptscriptfont1=\seveni
  \textfont2=\twelvesy   \scriptfont2=\tensy   \scriptscriptfont2=\sevensy
  \textfont3=\twelveex   \scriptfont3=\twelveex  \scriptscriptfont3=\twelveex
  \textfont\itfam=\twelveit
  \textfont\slfam=\twelvesl
  \textfont\bffam=\twelvebf \scriptfont\bffam=\tenbf
  \scriptscriptfont\bffam=\sevenbf
  \normalbaselines\rm}

%       tenpoint

%%
%%      Various internal macros
%%

\def\beginlinemode{\endmode
  \begingroup\parskip=0pt \obeylines\def\\{\par}\def\endmode{\par\endgroup}}
\def\beginparmode{\endmode
  \begingroup \def\endmode{\par\endgroup}}
\let\endmode=\par
{\obeylines\gdef\
{}}
\def\singlespace{\baselineskip=\normalbaselineskip}

\def\oneandahalfspace{\baselineskip=\normalbaselineskip
  \multiply\baselineskip by 3 \divide\baselineskip by 2}
\def\doublespace{\baselineskip=\normalbaselineskip \multiply\baselineskip by 2}

\newcount\firstpageno
\firstpageno=2
\footline={\ifnum\pageno<\firstpageno{\hfil}%
\else{\hfil\twelverm\folio\hfil}\fi}
\let\rawfootnote=\footnote              % We must set the footnote style
\def\footnote#1#2{{\rm\singlespace\parindent=0pt\rawfootnote{#1}{#2}}}
\def\raggedcenter{\leftskip=4em plus 12em \rightskip=\leftskip
  \parindent=0pt \parfillskip=0pt \spaceskip=.3333em \xspaceskip=.5em
  \pretolerance=9999 \tolerance=9999
  \hyphenpenalty=9999 \exhyphenpenalty=9999 }
\def\dateline{\rightline{\ifcase\month\or
  January\or February\or March\or April\or May\or June\or
  July\or August\or September\or October\or November\or December\fi
  \space\number\year}}
\def\received{\vskip 3pt plus 0.2fill
 \centerline{\sl (Received\space\ifcase\month\or
  January\or February\or March\or April\or May\or June\or
  July\or August\or September\or October\or November\or December\fi
  \qquad, \number\year)}}

%%
%%      Page layout, margins, font and spacing (feel free to change)
%%

\hsize=6.5truein
%\hoffset=1truein
\vsize=8.9truein
%\voffset=1truein
\parskip=\medskipamount
\twelvepoint            % selects twelvepoint fonts (cf. \tenpoint)
\doublespace            % selects double spacing for main part of paper (cf.
                        %       \singlespace, \oneandahalfspace)
\overfullrule=0pt       % delete the nasty little black boxes for overfull box

%%
%%      The user definitions for major parts of a paper (feel free to change)
%%

    % Preprint number at upper right of title page

\def\title                      %  Title on title page
  {\null\vskip 3pt plus 0.2fill
   \beginlinemode \doublespace \raggedcenter \bf}

\def\author                     %  Author(s) name(s)  on title page
  {\vskip 3pt plus 0.2fill \beginlinemode
   \singlespace \raggedcenter}

\def\affil                      % Affiliations (can intermix with \author)
  {\vskip 3pt plus 0.1fill \beginlinemode
   \oneandahalfspace \raggedcenter \sl}

\def\abstract                   % Begin abstract
  {\vskip 3pt plus 0.3fill \beginparmode
   \doublespace \narrower ABSTRACT: }

\def\endtitlepage               % End title page, begin body of paper
  {\endpage                     %       This subsumes \body
   \body}

\def\body                       % Begin text body;  can be used to end
  {\beginparmode}               % \title, \author, \affil, \abstract,
                                % \reference, or \figurecaption modes

\def\subhead#1{                 % Subhead;  NOTE enclose the text in {}
  \vskip 0.25truein             % e.g., \subhead{A. History of the Problem}
  {\raggedcenter #1 \par}
   \nobreak\vskip 0.25truein\nobreak}

\def\refto#1{$|{#1}$}           % For references in text as superscript

\def\references                 % Begin references -- basic format is Phys Rev
  {\subhead{References}         % I.e., volume, page, year (space after commas)
   \beginparmode
   \frenchspacing \parindent=0pt \leftskip=1truecm
   \parskip=8pt plus 3pt \everypar{\hangindent=\parindent}}

\gdef\refis#1{\indent\hbox to 0pt{\hss#1.~}}    % Ref list numbers.

\gdef\journal#1, #2, #3, 1#4#5#6{               % Journal reference.  Comma set
    {\sl #1~}{\bf #2}, #3, (1#4#5#6)}           % off: name, vol, page, year

\def\refstylenp{                % Nucl Phys(or Phys Lett) ref style: V, Y, P
  \gdef\refto##1{ [##1]}                                % Reference in text []
  \gdef\refis##1{\indent\hbox to 0pt{\hss##1)~}}        % Ref list numbers)
  \gdef\journal##1, ##2, ##3, ##4 {                     % Journal reference
     {\sl ##1~}{\bf ##2~}(##3) ##4 }}

\def\refstyleprnp{              % Input like pr, output like np!!
  \gdef\refto##1{ [##1]}                                % Reference in text []
  \gdef\refis##1{\indent\hbox to 0pt{\hss##1)~}}        % Ref list numbers)
  \gdef\journal##1, ##2, ##3, 1##4##5##6{               % Journal reference
    {\sl ##1~}{\bf ##2~}(1##4##5##6) ##3}}

\def\endreferences{\body}

\def\figurecaptions             % Begin figure captions
  { \beginparmode
   \subhead{Figure Captions}
}

\def\endpage                    %  Eject a page
  {\vfill\eject}

\def\endpaper                   %  Ways to say goodbye
  {\endmode\vfill\supereject}

%%
%%      Various little user definitions
%%

\def\ref#1{Ref. #1}                     %       for inline references
\def\Ref#1{Ref. #1}                     %       ditto

          % For citation of equation numbers
        %       ditto
                     %       ditto
                     %       ditto
                   %       ditto
                   %       ditto
\def\frac#1#2{{\textstyle{#1 \over #2}}}

\def\sla{\raise.15ex\hbox{$/$}\kern-.57em}
\def\leaderfill{\leaders\hbox to 1em{\hss.\hss}\hfill}
\def\twiddle{\lower.9ex\rlap{$\kern-.1em\scriptstyle\sim$}}
\def\bigtwiddle{\lower1.ex\rlap{$\sim$}}
\def\gtwid{\mathrel{\raise.3ex\hbox{$>$\kern-.75em\lower1ex\hbox{$\sim$}}}}
\def\ltwid{\mathrel{\raise.3ex\hbox{$<$\kern-.75em\lower1ex\hbox{$\sim$}}}}
\def\square{\kern1pt\vbox{\hrule height 1.2pt\hbox{\vrule width 1.2pt\hskip 3pt
   \vbox{\vskip 6pt}\hskip 3pt\vrule width 0.6pt}\hrule height 0.6pt}\kern1pt}

\def\m@th{\mathsurround=0pt }
\def\leftrightarrowfill{$\m@th \mathord\leftarrow \mkern-6mu
 \cleaders\hbox{$\mkern-2mu \mathord- \mkern-2mu$}\hfill
 \mkern-6mu \mathord\rightarrow$}
\def\overleftrightarrow#1{\vbox{\ialign{##\crcr
     \leftrightarrowfill\crcr\noalign{\kern-1pt\nointerlineskip}
     $\hfil\displaystyle{#1}\hfil$\crcr}}}

%% *********** New stuff follows *******************

\font\titlefont=cmr10 scaled\magstep3

\def\martinstyletitle                      %  Title on title page
  {\null\vskip 3pt plus 0.2fill
   \beginlinemode \doublespace \raggedcenter \titlefont}

\font\twelvesc=cmcsc10 scaled 1200

\def\author                     %  Author(s) name(s)  on title page
  {\vskip 3pt plus 0.2fill \beginlinemode
   \singlespace \raggedcenter\twelvesc}

%%
%%      AmSTeX compatability definitions
%%
%%      To run a TeX file originally intended for AmSTeX, only small changes
%%      should be necessary (I hope).  Use the line \input jnl at the start.
%%      Remove the lines \input amstex, \documentstyle{itpjnl} at the
%%      beginning;  also remove all the page layout stuff (\parindent=1cm,
%%      \hsize=5.28125in etc.)  The page layout is now done automatically.
%%      Also OMIT the qualifier \magnification=1200 when you IMPRINT the
%%      .dvi file.  (\TagsOnRight is harmless, you can take it out or leave
%%      it in.)  I believe most AmSTeX will work with no change.  One problem
%%      is \footnote, which is a little different in that it now needs to
%%      have an explicit asterisk *  (or whatever) included, like this:
%%              \footnote*{Text winds up at bottom of page.}
%%      This is discussed on p. 116 of the TeXbook.  IGNORE the AmSTeX
%%      documentation (if you can call it that);  refer to the TeXbook.
%%
%%      Note that many commands in AmSTeX have their equivalents in the
%%      TeXbook, perhaps with different names and slightly differing
%%      usage. E.g., the old \align in AmSTeX is replaced by \eqalign
%%      (p. 190) and \aligntag is replaced by \eqalignno (p. 192).
%%      \align and \aligntag still work, but I recommend that you use
%%      \eqalign and \eqalignno in documents run under jnl.
%%
%%      See me if you have any problems  -- Doug.
%%

\def\heading                            % Heading
  {\vskip 0.5truein plus 0.1truein      % e.g., \heading I. NOTES \endheading
   \beginparmode \def\\{\par} \parskip=0pt \singlespace \raggedcenter}

\def\subheading                         % Subheading
  {\vskip 0.25truein plus 0.1truein     % e.g., \subheading{A. The Problem}
   \beginlinemode \singlespace \parskip=0pt \def\\{\par}\raggedcenter}

\def\tag#1$${\eqno(#1)$$}

\def\align#1$${\eqalign{#1}$$}

\def\aligntag#1$${\gdef\tag##1\\{&(##1)\cr}\eqalignno{#1\\}$$
  \gdef\tag##1$${\eqno(##1)$$}}

\def\endaligntag{}

\def\overset #1\to#2{{\mathop{#2}\limits^{#1}}}
\def\underset#1\to#2{{\let\next=#1\mathpalette\undersetpalette#2}}
\def\undersetpalette#1#2{\vtop{\baselineskip0pt
\ialign{$\mathsurround=0pt #1\hfil##\hfil$\crcr#2\crcr\next\crcr}}}

%%
%%      Various little user definitions
%%

\def\ref#1{Ref.~#1}                     %       for inline references
\def\Ref#1{Ref.~#1}                     %       ditto
\def\[#1]{[\cite{#1}]}
\def\cite#1{{#1}}
\def\(#1){(\call{#1})}
\def\call#1{{#1}}
\def\taghead#1{}
\def\frac#1#2{{#1 \over #2}}

\def\12{{1\over2}}

\def\sla{\raise.15ex\hbox{$/$}\kern-.57em}
\def\leaderfill{\leaders\hbox to 1em{\hss.\hss}\hfill}
\def\twiddle{\lower.9ex\rlap{$\kern-.1em\scriptstyle\sim$}}
\def\bigtwiddle{\lower1.ex\rlap{$\sim$}}
\def\gtwid{\mathrel{\raise.3ex\hbox{$>$\kern-.75em\lower1ex\hbox{$\sim$}}}}
\def\ltwid{\mathrel{\raise.3ex\hbox{$<$\kern-.75em\lower1ex\hbox{$\sim$}}}}
\def\square{\kern1pt\vbox{\hrule height 1.2pt\hbox{\vrule width 1.2pt\hskip 3pt
   \vbox{\vskip 6pt}\hskip 3pt\vrule width 0.6pt}\hrule height 0.6pt}\kern1pt}
\def\tdot#1{\mathord{\mathop{#1}\limits^{\kern2pt\ldots}}}

\def\pmb#1{\setbox0=\hbox{#1}%
  \kern-.025em\copy0\kern-\wd0
  \kern  .05em\copy0\kern-\wd0
  \kern-.025em\raise.0433em\box0 }

\catcode`@=11
\newcount\tagnumber\tagnumber=0

\immediate\newwrite\eqnfile
\newif\if@qnfile\@qnfilefalse
\def\write@qn#1{}
\def\writenew@qn#1{}
\def\w@rnwrite#1{\write@qn{#1}\message{#1}}
\def\@rrwrite#1{\write@qn{#1}\errmessage{#1}}

\def\taghead#1{\gdef\t@ghead{#1}\global\tagnumber=0}
\def\t@ghead{}

\expandafter\def\csname @qnnum-3\endcsname
  {{\t@ghead\advance\tagnumber by -3\relax\number\tagnumber}}
\expandafter\def\csname @qnnum-2\endcsname
  {{\t@ghead\advance\tagnumber by -2\relax\number\tagnumber}}
\expandafter\def\csname @qnnum-1\endcsname
  {{\t@ghead\advance\tagnumber by -1\relax\number\tagnumber}}
\expandafter\def\csname @qnnum0\endcsname
  {\t@ghead\number\tagnumber}
\expandafter\def\csname @qnnum+1\endcsname
  {{\t@ghead\advance\tagnumber by 1\relax\number\tagnumber}}
\expandafter\def\csname @qnnum+2\endcsname
  {{\t@ghead\advance\tagnumber by 2\relax\number\tagnumber}}
\expandafter\def\csname @qnnum+3\endcsname
  {{\t@ghead\advance\tagnumber by 3\relax\number\tagnumber}}

\def\equationfile{%
  \@qnfiletrue\immediate\openout\eqnfile=\jobname.eqn%
  \def\write@qn##1{\if@qnfile\immediate\write\eqnfile{##1}\fi}
  \def\writenew@qn##1{\if@qnfile\immediate\write\eqnfile
    {\noexpand\tag{##1} = (\t@ghead\number\tagnumber)}\fi}
}

\def\callall#1{\xdef#1##1{#1{\noexpand\call{##1}}}}
\def\call#1{\each@rg\callr@nge{#1}}

\def\each@rg#1#2{{\let\thecsname=#1\expandafter\first@rg#2,\end,}}
\def\first@rg#1,{\thecsname{#1}\apply@rg}
\def\apply@rg#1,{\ifx\end#1\let\next=\relax%
\else,\thecsname{#1}\let\next=\apply@rg\fi\next}

\def\callr@nge#1{\calldor@nge#1-\end-}
\def\callr@ngeat#1\end-{#1}
\def\calldor@nge#1-#2-{\ifx\end#2\@qneatspace#1 %
  \else\calll@@p{#1}{#2}\callr@ngeat\fi}
\def\calll@@p#1#2{\ifnum#1>#2{\@rrwrite{Equation range #1-#2\space is bad.}
\errhelp{If you call a series of equations by the notation M-N, then M and
N must be integers, and N must be greater than or equal to M.}}\else %
{\count0=#1\count1=#2\advance\count1 by1\relax\expandafter\@qncall\the\count0,%
  \loop\advance\count0 by1\relax%
    \ifnum\count0<\count1,\expandafter\@qncall\the\count0,%
  \repeat}\fi}

\def\@qneatspace#1#2 {\@qncall#1#2,}
\def\@qncall#1,{\ifunc@lled{#1}{\def\next{#1}\ifx\next\empty\else
  \w@rnwrite{Equation number \noexpand\(>>#1<<) has not been defined yet.}
  >>#1<<\fi}\else\csname @qnnum#1\endcsname\fi}

\let\eqnono=\eqno
\def\eqno(#1){\tag#1}
\def\tag#1$${\eqnono(\displayt@g#1 )$$}

\def\aligntag#1\endaligntag
  $${\gdef\tag##1\\{&(##1 )\cr}\eqalignno{#1\\}$$
  \gdef\tag##1$${\eqnono(\displayt@g##1 )$$}}

\def\eqalignno#1{\displ@y \tabskip\centering
  \halign to\displaywidth{\hfil$\displaystyle{##}$\tabskip\z@skip
    &$\displaystyle{{}##}$\hfil\tabskip\centering
    &\llap{$\displayt@gpar##$}\tabskip\z@skip\crcr
    #1\crcr}}

\def\displayt@gpar(#1){(\displayt@g#1 )}

\def\displayt@g#1 {\rm\ifunc@lled{#1}\global\advance\tagnumber by1
        {\def\next{#1}\ifx\next\empty\else\expandafter
        \xdef\csname @qnnum#1\endcsname{\t@ghead\number\tagnumber}\fi}%
  \writenew@qn{#1}\t@ghead\number\tagnumber\else
        {\edef\next{\t@ghead\number\tagnumber}%
        \expandafter\ifx\csname @qnnum#1\endcsname\next\else
        \w@rnwrite{Equation \noexpand\tag{#1} is a duplicate number.}\fi}%
  \csname @qnnum#1\endcsname\fi}

\def\ifunc@lled#1{\expandafter\ifx\csname @qnnum#1\endcsname\relax}

\let\@qnend=\end\gdef\end{\if@qnfile
\immediate\write16{Equation numbers written on []\jobname.EQN.}\fi\@qnend}

\catcode`@=12

\catcode`@=11
\newcount\r@fcount \r@fcount=0
\newcount\r@fcurr
\immediate\newwrite\reffile
\newif\ifr@ffile\r@ffilefalse
\def\w@rnwrite#1{\ifr@ffile\immediate\write\reffile{#1}\fi\message{#1}}

\def\writer@f#1>>{}
\def\referencefile{%			  Stuff to write .REF file
  \r@ffiletrue\immediate\openout\reffile=\jobname.ref%
  \def\writer@f##1>>{\ifr@ffile\immediate\write\reffile%
    {\noexpand\refis{##1} = \csname r@fnum##1\endcsname = %
     \expandafter\expandafter\expandafter\strip@t\expandafter%
     \meaning\csname r@ftext\csname r@fnum##1\endcsname\endcsname}\fi}%
  \def\strip@t##1>>{}}

\def\citeall#1{\xdef#1##1{#1{\noexpand\cite{##1}}}}
\def\cite#1{\each@rg\citer@nge{#1}}	% Variable No. of args, separated by

\def\each@rg#1#2{{\let\thecsname=#1\expandafter\first@rg#2,\end,}}
\def\first@rg#1,{\thecsname{#1}\apply@rg}	% each@ag is a general purpose
\def\apply@rg#1,{\ifx\end#1\let\next=\relax%	  variable no. of arg. macro.
\else,\thecsname{#1}\let\next=\apply@rg\fi\next}% args separated by commas

\def\citer@nge#1{\citedor@nge#1-\end-}	% Check for M-N range (M and N numbers)
\def\citer@ngeat#1\end-{#1}
\def\citedor@nge#1-#2-{\ifx\end#2\r@featspace#1 % Single argument
  \else\citel@@p{#1}{#2}\citer@ngeat\fi}	% M-N range of arguments
\def\citel@@p#1#2{\ifnum#1>#2{\errmessage{Reference range #1-#2\space is bad.}%
    \errhelp{If you cite a series of references by the notation M-N, then M and
    N must be integers, and N must be greater than or equal to M.}}\else%
 {\count0=#1\count1=#2\advance\count1 by1\relax\expandafter\r@fcite\the\count0,
  \loop\advance\count0 by1\relax%	  Loop from M to N
    \ifnum\count0<\count1,\expandafter\r@fcite\the\count0,%
  \repeat}\fi}

\def\r@featspace#1#2 {\r@fcite#1#2,}	% Eat spaces at beginning or end of arg
\def\r@fcite#1,{\ifuncit@d{#1}%		  Cite individual reference
    \newr@f{#1}%
    \expandafter\gdef\csname r@ftext\number\r@fcount\endcsname%
                     {\message{Reference #1 to be supplied.}%
                      \writer@f#1>>#1 to be supplied.\par}%
 \fi%
 \csname r@fnum#1\endcsname}
\def\ifuncit@d#1{\expandafter\ifx\csname r@fnum#1\endcsname\relax}%
\def\newr@f#1{\global\advance\r@fcount by1%
    \expandafter\xdef\csname r@fnum#1\endcsname{\number\r@fcount}}

\let\r@fis=\refis			% Save old \refis, redefine
\def\refis#1#2#3\par{\ifuncit@d{#1}%      Use two params #2 #3 to strip blank
   \newr@f{#1}%
   \w@rnwrite{Reference #1=\number\r@fcount\space is not cited up to now.}\fi%
  \expandafter\gdef\csname r@ftext\csname r@fnum#1\endcsname\endcsname%
  {\writer@f#1>>#2#3\par}}

\def\ignoreuncited{%   redefine \refis if ignoring uncited references
   \def\refis##1##2##3\par{\ifuncit@d{##1}%
    \else\expandafter\gdef\csname r@ftext\csname r@fnum##1\endcsname\endcsname%
     {\writer@f##1>>##2##3\par}\fi}}

\def\r@ferr{\endreferences\errmessage{I was expecting to see
\noexpand\endreferences before now;  I have inserted it here.}}
\let\r@ferences=\references
\def\references{\r@ferences\def\endmode{\r@ferr\par\endgroup}}

\let\endr@ferences=\endreferences
\def\endreferences{\r@fcurr=0%		  Save old \endreferences, redefine
  {\loop\ifnum\r@fcurr<\r@fcount%	  Loop over refnum and produce text
    \advance\r@fcurr by 1\relax\expandafter\r@fis\expandafter{\number\r@fcurr}%
    \csname r@ftext\number\r@fcurr\endcsname%
  \repeat}\gdef\r@ferr{}\endr@ferences}

% Save old \endpaper, redefine it to write parting message.

\let\r@fend=\endpaper\gdef\endpaper{\ifr@ffile
\immediate\write16{Cross References written on []\jobname.REF.}\fi\r@fend}

\catcode`@=12

\citeall\refto		% These macros will generate citations
\citeall\ref		%
\citeall\Ref		%

% Belgium Talk (1992) published in 1993 original in Tex
% References updated to 1993, put on gr-qc Dec. 1995
%\input jnl
%\input reforder
%\input referencefile
%\ignoreuncited

\centerline {\bf Quantum Origin of Noise and Fluctuations in Cosmology}
\vskip .7cm
\centerline{ B. L. Hu }
\centerline{\sl Department of Physics, University of Maryland,
College Park, MD 20742}
\vskip .7cm
\centerline{ Juan Pablo Paz}
\centerline{\sl Theoretical Astrophysics, MS B288, LANL, Los Alamos,
NM 87545}
\vskip .7cm
\centerline{ Yuhong Zhang}
\centerline{\sl Biophysics Lab, CBER, Food and Drug Adminstration,
8800 Rockville Pike, Bethesda, MD 20982}
\vskip 1cm
\centerline{(pp\# 93-55,  August, 1992)}
\vskip 1cm
\centerline{\it Invited talk delivered by B. L. Hu at the Conference on}
\centerline{ The Origin of Structure in the Universe}
\centerline{\it  Chateau du Pont d'Oye,  Belgium, April, 1992}
\centerline
{\it Edited by E. Gunzig and P. Nardonne (Plenum Press, New York, 1993)}
\vskip 2cm

\centerline{\bf Abstract}
\vskip .7cm
 We address two basic issues in the theory of galaxy formation from
fluctuations of quantum fields: 1) the nature and origin of noise and
fluctuations and 2) the conditions for using a classical stochastic equation
for their description. On the first issue, we derive the influence
functional for a $\lambda \phi^4 $ field in a zero-temperature bath in de
Sitter universe and obtain the correlator for the colored noises of vacuum
fluctuations.  This exemplifies a new mechanism we propose for colored noise
generation which can  act as seeds for galaxy formation with non-Gaussian
distributions. For the second issue,
we present a (functional) master equation for the inflaton field
in de Sitter universe. By examining the form of the noise kernel
we study the decoherence of the long-wavelength sector and the conditions
for it to behave classically.

%\vskip 2cm
%\noindent $^{\ast}$: bitnet addresses: hu@umdhep, paz@umdhep, zhang@umdhep

%\body
%\endtitlepage
\vfill
\eject

\noindent {\bf 1.~ Galaxy Formation from Quantum Fluctuations}

A standard mechanism for galaxy formation is the amplification of primordial
density fluctuations by the evolutionary dynamics of  spacetime
\refto {Lifshitz, Bardeen}.
In the lowest order approximation the gravitational perturbations
(scalar perturbations for matter density and tensor perturbations
for gravitational waves) obey linear equations of motion.
Their initial values and distributions are stipulated--
oftentimes assumed to be a white noise spectrum. In these theories,
fashionable in the sixties and seventies, the primordial fluctuations
are classical in nature. The Standard model of Friedmann-Lemaitre-
Robertson-Walker with power-law dependence (on cosmic time)
generates a density contrast which turns out to be too small to account
for the observed galaxy masses. The observed nearly scale-invariant spectrum
also does not find any easy explanation in this model \refto{Peebles, ZelNov}.

The inflationary cosmology of the eighties \refto{Guth, AlbSte, LindeInf}
is based on the dynamics of a quantum field $\phi$
undergoing a phase transition.  The exponential expansion
of the scale parameter $a(t)= a_0 \exp (Ht) $
gives a scale-invariant spectrum naturally.
This is one of the many attractive features of the inflationary universe,
particularly with regard to the galaxy formation problem.
The primordial fluctuations are quantum in nature.
They arise from the fluctuations of the quantum field
which induces inflation, sometimes called the inflaton.  The density contrast
$\delta \rho / \rho$ can be shown to be related to the fluctuations of
the scalar field $\Delta \phi$ approximately by \refto{GalForInf}
$$
{{\delta \rho} \over \rho} \approx
{ {H \Delta \phi} \over {<\dot \phi>}}     \eqno(1.1)
$$

\noindent Here $H= \dot a / a$ is the Hubble expansion rate, assumed to be a
constant for the de Sitter phase of the evolution,
and $< ~~ >$ denotes average over some spatial range.
For the density contrasts to be within $ 10^{-4}$ when the modes enter the
horizon the coupling constant in the Higgs field (e.g. a $\lambda \phi^4$
theory) in the standard models of unified theories
has to be exceedingly small ($ \lambda \sim 10^{-12} $).

The main features of the inflationary cosmology are determined by the dynamics
of different sectors of the normal modes of the scalar field in relation to
the exponential Hubble expansion of the background spacetime.
The scalar field $\Phi$ evolves according to the equation
$$
\ddot \Phi + 3H \dot \Phi + V' (\Phi) = 0                        \eqno(1.2)
$$

\noindent where the potential $V(\Phi)$ can take on a variety of forms.
A common form for the discussion of the generic behavior of old \refto{Guth}
and chaotic \refto{LindeChaos} inflation is the $\phi^4$ potential

$$
V(\Phi) = {1\over 2} m^2 \Phi^2 + {1 \over {4!}} \lambda \Phi^4  \eqno(1.3)
$$

\noindent For new inflation \refto{AlbSte, LindeInf} to work,
the potential has to possess a flat pleatau, as in
the Coleman-Weinberg form. Another commonly used potential
is the exponential form \refto{LucMat} .

Consider a scalar inflation field in a de Sitter space.
In this so-called `eternal inflation' stage
the horizon size $ l_h=H^{-1} $ is fixed. The physical wavelength $l$ of a
mode of the inflation field is $ l=p^{-1}=a(t)/k $ where $ k $ is
the wave number of that mode. As the scale factor increases
exponentially, the wavelengths of many modes can
grow larger than the horizon size. After the end of the
de Sitter phase, the universe begins to reheat and turns into a
radiation-dominated Friedmann universe with
power law expansion $ a(t)\sim t^n $. In this phase,
the horizon size expands
much faster than the physical wavelength. So some inflaton modes that
left the de Sitter horizon will later reenter the Friedman horizon, i.e., the
physical wavelength becomes shorter than the horizon size in this
radiation or matter dominated phase. The fluctuations of these long-wavelength
inflaton modes that had gone out and later come back into the horizon
play an important role in determining the large scale density fluctuations
of the early universe which later evolve to galaxies. A common assumption
is that these long wavelength inflaton modes behave classically while the other
short wavelength inflaton modes behave like quantum fluctuations
\refto{GuthPiInf}.
While this overall picture is generally accepted, a fully quantum mechanical
description
of  the evolution of the inflaton field and its fluctuations undergoing
phase transitions in the inflationary universe is still lacking.
%In particular, one needs a theory which can explain how the large wavelength
%modes become classical, i.e., how the classical spectrum of the density
%fluctuations arises from quantum fluctuations.

One suggestion made by Starobinsky \refto{Staro86}
and Bardeen and Bublik \refto{BarBub} in what is known as
`stochastic inflation' is to  split the inflation field into two parts
at every instant according to their physical wavelengths, i.e.,
$$
 \Phi (x)= \phi(x) + \psi(x).                                      \eqno(1.4)
$$

\noindent The first part $\phi$ (the `system field') consists of
field modes whose physical wavelengths are longer than the de Sitter
horizon size $p< \epsilon H$. The second part $\psi$ (the
`environment field') consists of field
modes whose physical wavelengths are shorter than the horizon size
$p> \epsilon H$.
At early times the modes in the system behave with little difference
from that in Minkowsky space. Here $\epsilon$ is a small parameter measuring
their deviation from the Minkowsky behavior.  Inflation continuously shifts
more and more modes of the environment field into the system after their
physical wavelengths exceed the de Sitter horizon size.

Starobinsky's model treats a free, massless, conformally-coupled field.
With $m=0$ and $\lambda = 0$ in (1.3), substitution of (1.4) into (1.2)
gives an equation of motion for the system field $\phi$
$$
\ddot \phi(t) +3H \dot \phi + V'(\phi) = \xi (t)                 \eqno(1.5)
$$
where
$$
<\xi(t)> = 0, ~~~~<\xi(t)\xi(t')> = \delta (t-t')
$$

The common belief is that the bath field contributes a white noise source
\refto{Staro86, BarBub, Rey}. With this assumption,
the system field equation is thus rendered into a classical Langevin equation
with a white noise source.
A Fokker-Planck equation can also be derived which depicts
the evolution of the probability distribution of the scalar field
$ P(\phi,t) $
\refto{Graziani}. Much recent effort is devoted to the solution
of this stochastic equation for a description of the inflationary
transition and galaxy formation
problems.

Note that two basic assumptions are made in  transforming a quantum
field theoretic problem to a classical stochastic mechanics problem as in
the stochastic inflation program:
1) The low frequency scalar field modes (the system) behave classically.
2) The high frequency quantum field modes (the bath) behave like a
white noise.
Most previous researchers seem to hold the view that
the first condition is obvious and the second condition is proven.
In our view neither case is clear. We think that the first proposal is
plausible, but the proof is non-trivial while
the second proposal is dubious and  a correct proof does not yet exist.
%\refto{HabMij}
Overall, a more rigorous treatment of
the relation of quantum and classical fluctuations, and the source and nature
of noise is needed before a sound foundation for this program
can be established.

%One needs to spell out the conditions for it to be true (how are the systems
%chosen, how does the result depend on the coupling between the system
%and the bath.

On the first issue one needs to consider the conditions upon which
quantum fluctuations evolve to be classical. It requires first an understanding
of quantum to classical transition, which involves the decoherence process
\refto{envdec} \refto{conhis}.
It also questions the conditions when a quantity (field or geometry)
can be effectively viewed as fluctuation rather than part of the background.
Both quantum field and statistical considerations are needed to give a clear
picture of the relation of quantum to classical and background to
fluctuations .
In particular one needs to consider the
decoherence of different histories of quantum fields in a given spacetime
dynamics (in the context of semiclassical cosmology), and, more thoroughly,
that of the histories of spacetimes themselves (in the context of quantum
cosmology) \refto{decQC}.
Some work has appeared in addressing this aspect of the problem in inflationary
cosmology \refto{BraLafMij}.
Our current research on inflationary cosmology is directed towards clarifying
these two issues. We are using different concepts and approaches
in quantum mechanics \refto{HuZhaUnc} and quantum kinetic theory
%\refto{CalHuFluct}
to explore the relation between quantum and classical fluctuations, and
applying
some of the techniques attempted earlier in quantum cosmology \refto{decQC}
to decoherence in inflationary cosmology, %\refto{HPR},
but we shall not discuss this issue here.
Our concern here is mainly with the second proposal,
although the two issues are related and the theoretical framework we use here
can be used to address both.

%We have doubts on the second proposal and find the usual proofs
%largely unacceptable. The problem involves the interaction
%of two sectors of a free field via a moving partition-- one side feeding
%modes into the other due to red-shifting of the overall system. The claim
%is that the feeding sector acts like a white noise source to the receiving
%sector.  This problem is in the nature of quamtum field theory with
%moving boundaries, but most of the existing proofs are amiss in this basic
%respect. The conjecture may still be valid, although we have doubts on that,
%but the proofs attempted so far are inadequate.

On the issue of noise, note that
for a free field the  inflaton modes do not interact with the bath
modes and they do not interact with one another. These field modes behave like
a collection of non-interacting free particles in an ideal gas. The
separation is like a sieving partition which moves in time.
It is obvious that adding or taking away some particles (modes) from the
system should not disturb the motion
of other particles in the system. But the system as a whole may lose or gain
energy through the exchange of particles with the environment, which itself
is depleting in content. The common claim of researchers in stochastic
inflation
is that the effect of this infusion of modes on the system is like
a noise source , in particular, a white noise source for free fields.

%Moreover, only for a free massless conformal field in a de Sitter space,
%and only under the above special (time-dependent) system-bath field splitting,
%will the bath field induced noise be a linearly coupled white noise, and the
%stochastic process is a Markoffian process. Otherwise, the noise would
%%generall
%y
%be non-linearly coupled color noise.

There are two problems with this view. Theoretically, a rigorous treatment of
this problem requires a quantum field theory of open systems, which,
contrary to what is commonly perceived and practised, is not
a straightforward matter. What constituents the system actually
changes in time as it is constantly enhanced by modes from the environment
and interacts with them.
Physically, if one  works in formalisms which deal only with
pure states, as has been done so far in most papers written on this topic,
it is difficult to understand how the concept of noise arise.
%can be accomodated.
Even if one forces in the identification of a noise source by splitting the
fields and averaging part of them one cannot find a corresponding
dissipation force. This is an unsatisfactory feature since
physically noise and dissipation should always appear together
according to the general fluctuation-dissipation relation.
(Some authors misconstrue the  red-shift term $ 3H\dot\phi $
in the Klein-Gordon equation as dissipation. It is a mistake).
A correct treatment should use a formalism which can encompass
the statistical nature of mixed states and the dynamics of reduced
density matrices as we shall show below.

%Also, as we already pointed out, there is no mode-mode coupling,
%so each field mode evolves independently. From this point of view, there
%should be no noise and dissipation at all for each individual mode without
%mode-mode coupling.

Here we seek a more basic approach to this issue which removes these two
drawbacks.
1) We adopt a fully field-theoretical treatment of non-equilibrium
quantum systems.
We use the influence functional formalism to treat the system-bath
interaction and show how noise arises from quantum fields when
one field (or a sector therein ) is coarse-grained,
and how its averaged effect on another field (or sector) is
described in a functional master equation, or
a functional Fokker-Planck- Wigner or Langevin equation.
We show how one can identify the nature of noise corresponding
to different baths and system- bath couplings.
2) We discuss the more realistic abeit more difficult case of an
interacting system field and
propose a different mechanism for the generation of
noise in the inflationary universe, vis., colored noise generation
from the nonlinear interaction of the inflation quantum field.
We take the usual $\lambda \phi^4$ potential assumed
in most inflation models as example, although the mechanism of colored
noise generation illustrated thus is generic in nature. The colored noise
source produced in this way
provides a natural mechanism for the generation of non-Gaussian spectrum of
density perturbations \refto{nonGaussian}%{SalBonBar, Ortolan, Hodges, Yi}.

{}From the general statistical physics point of view, the above issues
which pervade in the problems of inflationary cosmology and quantum cosmology
have their roots in problems of quantum open systems,  many of them
can be understood from simple examples in quantum mechanics.
\refto{HuTsukuba}. We have studied these problems in the context of
non-equilibrium statistical mechanics using the paradigm of quantum Brownian
motion \refto{HPZ1,HPZ2}. We refer the reader to these papers for details
and for a comparison with the present field-theoretical problem
for the discussion of the same issues.
In Sec. 2 we discuss the generation of colored noise from interacting
quantum fields in Minkowsky spacetime, assuming for simplicity two scalar
fields
with the full range of modes and a bi-quadratic form of coupling.
In Sec. 3 we discuss the corresponding problem in de Sitter spacetime.
Once the noise source is derived, one can then solve the Langevin equation
for the inflaton field, or the Fokker-Planck-Wigner equation
for the distribution function of the scalar field. We only write down the
master equation here.
In the discussion section (Sec. 4)
we summarize our findings,
and discuss how realistic our assumptions are,
and project possible problems in its consequences.
The main aim of this work, which is the first part of a project on noise,
fluctuations and structure formation, is to show how noise arises from
interacting quantum fields, or , more specifically,
in a fully quantum field-theoretical context,
how different noise sources (usually colored)
can arise from different  (nonlinear) interactions between the system and the
environment fields.
In the second part of this project currently under investigation, %\refto{HP},
we shall describe from the
stochastic dynamics of quantum fields in the early universe
how structures are formed from general fluctuations described by colored
noises.

%\vskip 2cm
\vfill
\eject

\noindent {\bf 2. ~~Colored Noise from Interacting Quantum Fields in Minkowski
Spacetime}

\vskip .5cm
We first consider quantum fields in a  Minkowski spacetime. The
separation of a single field into the high and low momentum sectors are
rather cumbersome to carry out, so for simpicity we will
consider two independent self-interacting scalar fields
$ \phi(x) $ depicting the system, and $ \psi(x) $ depicting the bath.
The physics is expected to be similar to the partitioned case.
The classical action for these two fields are given respectively by:
$$
S[\phi]
=\int d^4x~\Bigl\{
  {1\over 2}\partial_{\nu}\phi(x)\partial^{\nu}\phi(x)
 -{1\over 2}m^2_{\phi}\phi^2(x)
 -{1\over 4!}\lambda_{\phi}\phi^4(x) \Bigr\}
                                                        \eqno(2.1)
$$
$$
S[\psi]
=\int d^4x~\Bigl\{
  {1\over 2}\partial_{\mu}\psi(x)\partial^{\mu}\psi(x)
 -{1\over 2}m^2_{\psi}\psi^2(x)
 -{1\over 4!}\lambda_{\psi}\psi^4(x) \Bigr\}
=S_0[\psi]+S_I[\psi]
                                         \eqno(2.2)
$$

\noindent where $ m_{\phi} $ and $ m_{\psi} $ are the bare masses of
$ \phi(x) $ and $ \psi(x) $ fields respectively. Both fields have a
quartic self-interaction with the bare
coupling constants $ \lambda_{\phi} $ and $ \lambda_{\psi} $.
In (2.2) we have written $S[\psi]$ in terms of a free part $S_0$ and
an interacting part $S_I$ which contains $ \lambda_{\psi}$.
We assume that these two scalar fields interact via a bi-quadratic coupling
$$
S_{int}
=\int d^4x~\Bigl\{
 -\lambda_{\phi\psi}\phi^2(x)\psi^2(x) \Bigr\}
                                                        \eqno(2.3)
$$
\noindent and also that all three coupling constants
$ \lambda_{\phi} $, $ \lambda_{\psi} $ and $ \lambda_{\phi\psi} $
are small parameters of the same order.
The total classical action of the combined system plus bath field is
then given by
$$
S[\phi,\psi]
=S[\phi]+S[\psi]+S_{int}[\phi,\psi]
                                                        \eqno(2.4)
$$

\noindent The total density matrix of the combined system plus bath field is
defined by
$$
\rho[\phi,\psi,\phi',\psi',t]
=<\phi,\psi|~\hat\rho(t)~|\phi',\psi'>
                                                        \eqno(2.5)
$$
\noindent where $ |\phi> $ and $ |\psi> $ are the eigenstates of the field
operators $ \hat\phi(x) $ and $ \hat\psi(x) $, namely,
$$
\hat\phi(\vec x) |\phi>=\phi(\vec x) |\phi>, ~~~
\hat\psi(\vec x) |\psi> = \psi(\vec x) |\psi>
                                                        \eqno(2.6)
$$
%\noindent Its time evolution is given by
%$$
%\eqalign{
%& \rho[\phi_f,\psi_f;\phi'_f,\psi'_f,t] \cr
%& ~~~~~~~
%  =\int d\phi_i(\vec x)\int d\psi_i(\vec x)~
%   J[\phi_f,\psi_f,\phi'_f,\psi'_f,t~
%     |~\phi_i,\psi_i,\phi'_i,\psi'_f,t_0] \cr
%& ~~~~~~~ \times
%   \rho[\phi_i,\psi_i;\phi'_i,\psi'_i;t_0] \cr }
%                                                        \eqno(2.8)
%$$
%
%\noindent where $J$ is the propagator of the total density matrix
%
%$$
%\eqalign{
%& J[\phi_f,\psi_f,\phi'_f,\psi'_f,t~
%    |~\phi_i,\psi_i,\phi'_i,\psi'_f,t_0] \cr
%& ~~~~~~~
%  =\int\limits_{\phi_i(\vec x)}^{\phi_f(\vec x)}D\phi
%   \int\limits_{\psi_i(\vec x)}^{\psi_f(\vec x)}D\psi
%   \int\limits_{\phi'_i(\vec x)}^{\phi'_f(\vec x)}D\phi'
%   \int\limits_{\psi'_i(\vec x)}^{\psi'_f(\vec x)}D\psi'~
%   \exp i\Bigl\{ S[\phi,\psi]-S[\phi',\psi'] \Bigr\} \cr }
%                                                        \eqno(2.9)
%$$

Since we are primarily interested in the behavior of the system, and of
the environment only to the extent in
how it influences the system, the quantity of
relevance is the reduced density matrix defined by
$$
\rho_{red}[\phi,\phi',t]=\int d\psi \rho[\phi,\psi,\phi',\psi,t]
                                                           \eqno(2.7)
$$
%\noindent The integral $~\int d\phi(\vec s)~$ in (2.7) denotes the functional
%integral over the Hilbert space of all quantum states of the $ \phi $ field,
%while $~\int D\phi~$ denotes the functional path integral over all possible
%histories of the $ \phi $ field under some boundary conditions.
%(Here $ \hbar=1 $.)
%
For technical convenience, let us assume that the total density matrix
at an initial time is factorized, i.e.,
that the system and bath are statistically independent,
$$
\hat\rho(t_0)
=\hat\rho_{\phi}(t_0)\times\hat\rho_{\psi}(t_0)
                                                        \eqno(2.8)
$$
\noindent where $ \hat\rho_r(t_0) $ and $ \hat\rho_{\psi}(t_0) $ are the
initial
density matrix operator of the $ \phi $ and $ \psi $ field respectively,
the former being equal to the reduced density matrix $\hat \rho_r$ at $t_0$
by this assumption.
The reduced density matrix of the system field $ \phi(x) $ evolves
in time following
$$
\rho_r[\phi_f,\phi'_f,t]
=\int d\phi_i\int d\phi'_i~
 J_r[\phi_f,\phi'_f,t~|~\phi_i,\phi'_i,t_0]~
 \rho_r[\phi_i,\phi'_i,t_0]
                                                        \eqno(2.9)
$$

\noindent where $J_r$ is the propagator of the reduced density matrix:
$$
J_r[\phi_f,\phi'_f,t~|~\phi_i,\phi'_i,t_0]
=\int\limits_{\phi_i(\vec x)}^{\phi_f(\vec x)}D\phi
 \int\limits_{\phi'_i(\vec x)}^{\phi'_f(\vec x)}D\phi'~
 \exp i\Bigl\{ S[\phi]-S[\phi'] \Bigr\}~
  F[\phi,\phi']
                                                        \eqno(2.10)
$$

\noindent The influence functional $ ~F[\phi,\phi']~ $ is defined as
$$
\eqalign{
F[\phi,\phi']=
&  \int d\psi_f(\vec x)
   \int d\psi_i(\vec x)
   \int d\psi'_i(\vec x)~
   \rho_{\psi}[\psi_i,\psi'_i,t_0]~
   \int\limits_{\psi_i(\vec x)}^{\psi_f(\vec x)}D\psi
   \int\limits_{\psi'_i(\vec x)}^{\psi_f(\vec x)}D\psi' \cr
&  \times\exp i\Bigl\{S[\psi]+S_{int}[\phi,\psi]
    -S[\psi']-S_{int}[\phi',\psi'] \Bigr\} \cr }
                                                        \eqno(2.11)
$$

\noindent which summarizes the averaged effect of the bath on the system.
The influence action $ \delta A[\phi,\phi'] $
and the effective action $ A[\phi,\phi'] $ are defined as
$$
F[\phi,\phi']=\exp i\delta A[\phi,\phi']
                                                        \eqno(2.12)
$$
$$
A[\phi,\phi']
=S[\phi]-S[\phi']+\delta A[\phi,\phi']
                                                        \eqno(2.13)
$$

The above is the formal framework we shall adopt. Let us now begin the
technical
discussion of how to evaluate the influence action perturbatively. If
$ \lambda_{\phi\psi} $ and $ \lambda_{\psi} $ are assumed to be small
parameters,
%the self-interaction and the mutual coupling parts of the actions
%$ S_I[\psi] $ and $ S_{int}[\phi,\psi] $ in (2.18) and (2.3)
%are small perturbation of the same order.
the influence functional can be calculated perturbatively
by making a power expansion of $ \exp i\bigl\{S_{int}+S_I\bigr\} $.
Up to the second order in $~\lambda $,
%i.e., $ ~\lambda^2_{\psi},~\lambda^2_{\phi\psi}~ $ or
%$ ~\lambda_{\psi}\lambda_{\phi\psi}~ $
and first order in $\hbar$ (one-loop), the influence action is given by
$$
\eqalign{
\delta A[\phi,\phi']
& =~~\biggl\{
   <S_{int}[\phi,\psi]>_0
  -<S_{int}[\phi',\psi']>_0 \biggr\} \cr
& +{i\over 2}\biggl\{
   <\Bigl[S_{int}[\phi,\psi]\Bigr]^2>_0
  -\Bigl[<S_{int}[\phi,\psi]>_0\Bigr]^2 \biggr\} \cr
& -~i\biggl\{<S_{int}[\phi,\psi]S_{int}[\phi',\psi']>_0
   -<S_{int}[\phi,\psi]>_0
    <S_{int}[\phi',\psi']>_0 \biggr\} \cr
& +{i\over 2}\biggl\{
    <\Bigl[S_{int}[\phi',\psi']\Bigr]^2>_0
   -\Bigl[<S_{int}[\phi',\psi']>_0\Bigr]^2 \biggr\} \cr }
                                                        \eqno(2.14)
$$
\noindent where the quantum average of a physical variable $Q[\psi, \psi']$
over the unperturbed action $ S_0[\psi] $ is defined by
$$
\eqalign{
<Q[\psi,\psi']>_0
& =\int d\psi_f(\vec x)\int d\psi_i(\vec x)\int d\psi'_i(\vec x)~
    \rho_{\psi}[\psi_i,\psi'_i,0]  \cr
& \times
    \int\limits_{\psi_i(\vec x)}^{\psi_f(\vec x)} D\psi
    \int\limits_{\psi'_i(\vec x)}^{\psi_f(\vec x)} D\psi'~
    \exp i\Bigl\{ S_0[\psi]-S_0[\psi'] \Bigr\}
     \times Q[\psi,\psi'] \cr
& \equiv Q\Bigl[{\partial\over i\partial J_1(x)},~
    -{\partial\over i\partial J_2(x)} \Bigr]~
    F^{(1)}[J_1,J_2]~\biggl|_{J_1=J_2=0} \cr }
                                                        \eqno(2.15)
$$

\noindent Here, $ F^{(1)}[J_1,J_2] $ is the
influence functional of the free bath field, assuming a linear coupling with
external sources $J_1$ and $J_2$.

$$
\eqalign{
F^{(1)}[J_1,J_2]
&=\int d\psi_f(\vec x)
  \int d\psi_i(\vec x)
  \int d\psi'_i(\vec x)~
  \rho_{\psi}[\psi_i,\psi'_i,t_0]
  \int\limits_{\psi_i(\vec x)}^{\psi_f(\vec x)} D\psi
  \int\limits_{\psi'_i(\vec x)}^{\psi_f(\vec x)} D\psi' \cr
& \times\exp i\Bigl\{ S_0[\psi]+\int d^4x J_1(x)\psi(x)
    -S_0[\psi']-\int d^4x J_2(x)\psi'(x) \Bigr\} \cr }
                                                        \eqno(2.16)
$$

\noindent Let us define the following free propagators of the $ \psi $ field
$$
<\psi(x)\psi(y)>_0=iG_{++}(x,y)
                                                        \eqno(2.17)
$$
$$
<\psi'(x)\psi'(y)>_0=-iG_{--}(x,y)
                                                        \eqno(2.18)
$$
$$
<\psi(x)\psi'(y)>_0=-iG_{+-}(x,y)
                                                        \eqno(2.19)
$$
\noindent Then the influence action is given by

$$
\eqalign{
\delta A[\phi,\phi']
& =\int d^4x\Bigl\{-\lambda_{\phi\psi}
   iG_{++}(x,x)\phi^2(x)\Bigl\} \cr
& -\int d^4x\Bigl\{-\lambda_{\phi\psi}
   iG_{++}(x,x)\phi'^2(x)\Bigr\} \cr
& +\int d^4x\int d^4y~\lambda^2_{\phi\psi}\phi^2(x)~
   \Bigl\{-iG^2_{++}(x,y)\Bigr\}~\phi^2(y) \cr
& -2\int d^4x\int d^4y~\lambda^2_{\phi\psi}\phi^2(x)~
   \Bigl\{-iG^2_{+-}(x,y)\Bigr\}~\phi'^2(y) \cr
& +\int d^4x\int d^4y\lambda^2_{\phi\psi}\phi'^2(x)~
   \Bigl[-iG^2_{--}(x,y)\Bigr\}~\phi'^2(y) \cr }
                                                        \eqno(2.20)
$$

Note that if the bath is at zero temperature, i.e.,
if the bath field $ \psi $ is in a vacuum state,
$$
\hat\rho_b(t_0)=|0><0|                                   \eqno(2.21)
$$
\noindent then the influence functional (2.16) is the so-called Schwinger-
Keldysh or closed-time-path (CTP) or `in-in' vacuum generating functional
\refto{ctp} %{SchKel, ChineseCTP, DeWJor, CH87},
and the influence action (2.20) is the usual CTP or
in-in vacuum effective action. In such cases, the propagators
(2.17)-(2.19)
are just the well known Feynman, Dyson and positive-frequency
Wightman propagators of a free scalar field given respectively by,
$$
G_{++}(x,y)=G_F(x-y)
=\int{d^np\over (2\pi)^n}e^{ip(x-y)}
 {1\over p^2-m^2_{\psi}+i\epsilon}
                                                        \eqno(2.22)
$$
$$
G_{--}(x,y)=G_D(x-y)
=\int{d^np\over (2\pi)^n}e^{ip(x-y)}
 {1\over p^2-m^2_{\psi}-i\epsilon}
                                                        \eqno(2.23)
$$
$$
G_{+-}(x,y)=G^+(x-y)
=\int{d^np\over (2\pi)^2}e^{ip(x-y)}2\pi
i\delta(p^2-m^2_{\psi})\theta(p^0)
                                                        \eqno(2.24)
$$

The perturbation calculation for $\lambda \phi^4$ theory in the CTP formalism
has been carried out before for quantum fluctuations \refto{CH87}
and for coarsed-grained fields \refto{cgea, SinHu}.
%Since the perturbation calculation and renormalization procedure of the
%bi-quadratic interaction $ S_{int}[\phi,\psi] $ here is very similar to
%that of the $ \lambda\psi^4 $ field theory, we will only give the results
%here and refer the reader to the above references for details.
We find the effective action for this biquadratically-coupled
system-bath scalar field model to be
$$
\eqalign{
A[\phi,\phi']
& =\Bigl\{S[\phi]+\delta S_1[\phi]+\delta_2[\phi]\Bigr\}
  -\Bigl\{S[\phi']+\delta S_1[\phi']+\delta_2[\phi']\Bigr\}
  +\delta A[\phi,\phi'] \cr
& =S_{ren}[\phi]+\int d^4x\int d^4y~{1\over 2}
   \lambda^2_{\phi\psi}\phi^2(x)V(x-y)\phi^2(y) \cr
& -S_{ren}[\phi']-\int d^4x\int d^4y~{1\over 2}
   \lambda^2_{\phi\psi}\phi'^2(x)V(x-y)\phi'^2(y)\cr
& -\int\limits_{t_0}^tds_x\int d^3\vec x
   \int\limits_{t_0}^{s_y}ds_y\int d^3\vec y~
   \lambda^2_{\phi\psi}
   \Bigl[\phi^2(x)-\phi'^2(x)\Bigr] \cr
&  ~~~~~ \times\eta(x-x')
   \Bigl[\phi^2(y)+\phi'^2(y)\Bigr] \cr
& +i\int\limits_{t_0}^tds_x\int d^3\vec x
   \int\limits_{t_0}^{s_x}ds_y\int d^3\vec y~
   \lambda^2_{\phi\psi}
   \Bigl[\phi^2(x)-\phi'^2(x)\Bigr] \cr
&  ~~~~~ \times \nu(x-y)
   \Bigl[\phi^2(y)-\phi'^2(y)\Bigr] \cr }
                                                        \eqno(2.25)
$$

\noindent Here $ S_{ren}[\phi] $ is the renormalized action of the $\phi $
field,
now with physical mass $ m^2_{\phi r} $ and physical coupling constant
$ \lambda_{\phi r} $, namely,
$$
S_{ren}[\phi]
=\int d^4x\Bigl\{{1\over 2}\partial_{\mu}\phi\partial^{\mu}\phi
-{1\over 2}m^2_{\phi r}\phi^2-{1\over 4!}\lambda_{\phi r}\phi^4\Bigr\}
                                                        \eqno(2.26)
$$

\noindent and the kernel for the non-local potential in (2.25) is
$$
V(x-y)
=\mu(x-y)-sgn(s_x-s_y)\eta(x-y)
                                                        \eqno(2.27)
$$
\noindent which is symmetric.

Here $\eta$ and $\nu$ and $\mu$ are real nonlocal kernels
$$
\eta(x-y)
={1\over 16\pi^2}\int {d^4p\over (2\pi)^4}~e^{ip(x-y)}
{}~\pi\sqrt{1-{4m^2_{\psi}\over p^2}}~\theta(p^2-4m^2_{\psi})
 \times isgn(p_0)
                                                        \eqno(2.28)
$$
$$
\nu(x-y)
={2\over 16\pi^2}\int{d^4p\over (2\pi)^4}~e^{ip(x-y)}
{}~\pi~\sqrt{1-{4m^2_{\psi}\over p^2}}~\theta(p^2-4m^2_{\psi})
                                                        \eqno(2.29)
$$
$$
\mu(x-y)
=-{2\over 16\pi^2}\int{d^4p\over (2\pi)^4}~e^{ip(x-y)}
 \int\limits_0^1d\alpha\ln\Bigl|1-i\epsilon-
 \alpha(1-\alpha){p^2\over m^2_{\psi}}\Bigr|
                                                        \eqno(2.30)
$$

The imaginary part of the influence functional
can be viewed as arising from a noise source $ \xi(x)$ whose
distribution functional is given by
$$
P[\xi]
=N\times\exp\biggl\{-{1\over 2}\int d^4x \int d^4y
 \xi^2(x)~\lambda^{-2}_{\phi\psi}\nu^{-1}(x-y)~\xi^2(y)\biggr\}
                                                        \eqno(2.31)
$$

\noindent  where $ N $ is a normalization constant. The action describing
the noise $ \xi(x) $ and system field $ \phi(x) $ coupling is
$$
\int d^4x~\Bigl\{~\xi(x) \phi^2(x)~\Bigr\}
                                                        \eqno(2.32)
$$

\noindent
%As we can see this nonlinearly-coupled colored noise arises from
%a biquadratic coupling between two interacting quantum fields.
In the associated functional Langevin equation for the field, the
corresponding stochastic force arising from the biquadratic coupling
we have assumed is
$$
F_{\xi}(x)\sim\xi(x)\phi(x)
                                                        \eqno(2.33)
$$

\noindent which constitues a multiplicative noise \refto{HPZ2}.
%[See, e.g. \ref{stochastic}].

{}From the influence action (2.25), it is seen that the dissipation generated
in the system by this noise is of the nonlinear non-local type.
If we define the dissipation kernel $ \gamma(x-y) $ by
$$
\eta(x-y)
={\partial\over\partial (s_x-s_y)}\gamma(x-y)
                                                        \eqno(2.34)
$$

\noindent then
$$
\gamma(x-y)
={1\over 16\pi^2}\int {d^4p\over (2\pi)^4}~e^{ip(x-y)}
 \pi\sqrt{1-{4m^2_{\psi}\over p^2}}~
 \theta(p^2-4m^2_{\psi})~{1\over |p_0|}
                                                        \eqno(2.35)
$$

\noindent In the Langevin field equation, the dissipative force is

$$
F_{\gamma}(x)\sim
\Biggl\{\int d^4y~\eta(x-y)\phi^2(y)~\Biggr\}\phi(x)
                                                        \eqno(2.36)
$$

As discussed in  \ref{HPZ2},
we find that a fluctuation-dissipation relation exists between
the dissipation kernel (2.34) and the noise kernel (2.29) :

$$
\nu(x)=\int d^4y~K(x-y)\eta(y)
                                                        \eqno(2.37)
$$

\noindent where
$$
\eqalign{ K(x-y)
& =\delta^3(\vec x-\vec y)
   \int\limits_{-\infty}^{+\infty}
   {d\omega\over 2\pi}~
    e^{i\omega(s_x-s'_x)}|\omega| \cr
& =\delta^3(\vec x-\vec y)
   \int\limits_0^{+\infty}{d\omega\over\pi}
   \omega\cos\omega(s_x-s_y) \cr }
                                                        \eqno(2.38)
$$

\noindent Apart from the delta function $ \delta^3(\vec x-\vec x') $, the
convolution kernel for quantum fields has exactly the same form as for
the quantum Brownian harmonic oscillator with linear  or nonlinear
dissipations at zero temperature.
%This reaffirms our earlier claim
%that the  fluctuation-dissipation relation is a categorical relation.
%\refto{HPZ2}.

Thus we have given an explicit first-principle derivation of noise from
quantum fluctuations of interacting quantum fields.
We want to make three comments before closing this section.
First, note that here, as distinct from the free field case of \ref{Staro86},
the noise arises only because the coupling $\lambda_ {\phi\psi}$
between the system and the environment field is non-zero.
Second, it would be of interest to find the conditions upon which the
colored noise appears as white, i.e., $\nu(s) \rightarrow \delta(s)$
independent
of the detailed form of nonlinear coupling. This is possible from the
quantum mechanical cases studied in \ref{HPZ2}.
It could be at high temperature, or by a proper choice of the form of
the spectral density of the bath. But in field theory the second alternative
is not obviously implementable.
Third, we have discussed a zero-temperature bath here, where the noise
is of purely quantum nature, i. e., arising from vacuum fluctuations.
One can easily include finite temperature baths and deduce the noise from
thermal fluctuations of the bath. This is similar to the attempts of
\ref{CorBru}. Noises in finite temperature fields are discussed in
\ref{Zhang} for both Minkowski and de Sitter spacetimes.

%\vskip 0.5in
\vfill
\eject

\noindent {\bf 3.~ Master Equation with Colored Noise in de Sitter Universe}

\vskip 0.2in

We shall now proceed to calculate the influence functional for
an interacting field in de Sitter universe and identify the noise source.
Following \ref{Zhang} we shall derive the master equation from this
influence functional
for a special case and use it to examine the issue of decoherence.
This equation and its associated Langevin or
Fokker-Planck equation would enable one to calculate the fluctuation
spectrum as a classical stochastic dynamics problem.

Consider a real, gauge singlet, massive, $ \lambda\Phi^4 $
self-interacting scalar field in a de Sitter
spacetime with metric
$$
ds^2=g_{\mu\nu}dx^{\mu}dx^{\nu}=dt^2-a^2(t)d\vec x^2           \eqno(3.1)
$$
\noindent In the inflationary regime of interest,
the scalar factor $ a(t) $ expands exponentially in cosmic time $ t $
$$
a(t)=a_0\exp Ht
                                                        \eqno(3.2)
$$

\noindent The classical action of the inflaton field $ \Phi(x) $ is

$$
S[\Phi]=S_0[\Phi]+S_I[\Phi]
                                                        \eqno(3.3)
$$

\noindent where

$$
S_0[\Phi]
=\int d^nx\sqrt{-g(x)}\Bigl\{
 {1\over 2}g^{\mu\nu}\partial_{\mu}\Phi\partial_{\nu}\Phi
+{1\over 2} \xi_n R(t)\Phi^2 \Bigr\}
                                                        \eqno(3.4)
$$

\noindent is that part of the classical action describing a free, massless,
conformally coupled scalar field, and

$$
S_I[\Phi]
=\int d^nx\sqrt{-g(x)}\Bigl\{
 +{1\over 2}m^2\Phi^2
 +{1\over 2} \xi_n \xi R(x)\Phi^2
 -{1\over 4!}\lambda\Phi^4 \Bigr\}
                                                        \eqno(3.5)
$$

\noindent is the remaining (interactive) terms with contributions
from nonzero $m, \lambda$, and $\xi$, i.e., massive, self-interacting, or
non-conformal coupling. Here we use $\xi=0$ for conformal
coupling and $\xi=1$ for minimal coupling in four dimensions and
$ \xi_n = {(n-2)\over 4(n-1)}$ is a constant which is equal to 1/6
in 4-dimensions.

In the above,

$$
R(t)={6\over a^2(t)}{\ddot a(t)\over a(t)}
                                                        \eqno(3.6)
$$

\noindent is the scalar curvature,  and
$ ~\sqrt{-g(x)}=a^{n-1}(t),~ $.

In the Starobinsky scheme, one makes a system-bath field splitting

$$
\Phi(\vec x,t)=\phi(\vec x,t)+\psi(\vec x,t)
                                                        \eqno(3.7)
$$

\noindent such that the system field is defined by

$$
\phi(\vec x,t)
=\int\limits_{|\vec k|<\Lambda}{d^3\vec k\over (2\pi)^3}~
 \Phi(\vec k,t)~\exp ip\cdot x
                                                        \eqno(3.8)
$$

\noindent and the bath field is defined by
$$
\psi(\vec x,t)
=\int\limits_{|\vec k|>\Lambda}{d^3\vec k\over (2\pi)^3}~
 \Phi(\vec k,t)~\exp ip\cdot x
                                                        \eqno(3.9)
$$

\noindent where $ \Lambda $ is the cutoff wavenumber determined by the horizon
size. The  system field $ \phi(x) $ contains the long wavelength modes, which
undergoes a slow-rollover phase transition in the inflation period,
while the bath field $ \psi $ contains the short wavelength modes,
which are the quantum fluctuations.
With this splitting, the classical action (3.3) can be written as
$$
S[\Phi]=S[\phi]+S[\psi]+S_{int}[\phi,\psi]
                                                        \eqno(3.10)
$$
\noindent where
$$
S_{int}[\phi,\psi]
=\int d^nx \sqrt{-g(x)}\Bigl\{
 -{1\over 6}\lambda\phi^3\psi
 -{1\over 4}\lambda\phi^2\psi^2
 -{1\over 6}\lambda\phi\psi^3 \Bigr\}
                                                        \eqno(3.11)
$$

One can define the influence functional in de Sitter space
in a similar way as in the Minkowski space.
%Namely, the reduced density matrix of the system field $ \phi $ in de
%Sitter space and its propagator can be defined exactly as in (2.11)
%and (2.12), the influence functional and the influence action of $ \phi $
%field in de Sitter space are defined also similar to (2.13) and
%(2.14).
Assuming that  $ S_I[\psi] $ and $ S_{int}[\phi,\psi] $ are small
perturbations,
Then up to the second order in $ \lambda $ and to the first loop, one can show
that the influence action of the system field is given by
%
%DO WE WANT 4 or n dimensions in the integrals below?
%
$$
\eqalign{
& \delta A[\phi,\phi'] \cr
& =\int d^nx~a^{n-1}(t)~\Bigl\{
    -{1\over 2}\lambda<\psi^2(x)>_0\phi^2(x)
    +{1\over 2}\lambda<\psi'^2(x)>_0\phi'^2(x)\Bigr\} \cr
& +\int d^nx~a^{n-1}(t) \int d^nx'~a^{n-1}(t')~
    {i\over 4}\lambda M^2(x) \phi^2(x') \cr
&    ~~~~ \times\biggl\{
    \Bigl[<\psi(x)\psi(x')>_0\Bigr]^2-
    \Bigl[<\psi(x)\psi'(x')>_0\Bigr]^2\biggr\} \cr
& +\int d^nx~a^{n-1}(t) \int d^nx'~a^{n-1}(t')~
    {i\over 4}\lambda M^2(x) \phi'^2(x') \cr
&    ~~~~ \times\biggl\{
    \Bigl[<\psi'(x)\psi'(x')>_0\Bigr]^2-
    \Bigl[<\psi'(x)\psi(x')>_0\Bigr]^2\biggr\} \cr
& +\int d^nx~a^{n-1}(t) \int d^nx'~a^{n-1}(t')~
    {i\over 16}\lambda^2\phi^2(x)\Bigl[<\psi(x)
    \psi(x')>_0\Bigr]^2\phi^2(x) \cr
& -2\int d^nx~a^{n-1}(t) \int d^nx'~a^{n-1}(t')~
    {i\over 16}\lambda^2\phi^2(x)\Bigl[<\psi(x)
    \psi'(x')>_0\Bigr]^2\phi'^2(x') \cr
& +\int d^nx~a^{n-1}(t) \int d^nx'a^{n-1}(t')~
    {i\over 16}\lambda^2\phi'^2(x)\Big[<\psi'(x)
    \psi'(x')>_0\Bigr]^2\phi'^2(x') \cr }
                                                        \eqno(3.12)
$$

\noindent where $M^2(x) = m^2+\xi_n \xi R(x) $ and
the quantum average over a conformally-coupled massless free field
$ <~~~>_0 $ is defined similar to (2.17-19).
Note at the one loop level, (3.12) is similar to (2.20),
%CHANGE EQNO
apart from the mass coupling and the non-conformal coupling terms
in the original classical action (3.5).  The other two
interaction terms in (3.11) do not contribute at the one loop level.

Since the bath field $ \psi(\vec x) $ only contains high momentum
(long wavelength) modes, when we calculate the Feynman diagrams of (3.12)
by dimensional regularization, the momentum space integrations of Feynman
diagrams are restricted to the region which is outside of the sphere with
radius
$\Lambda$. This incomplete integration region in momentum space creates
some technical difficulty. For simplicity, we
extend the range of all integrations in the Feynman diagrams to cover the whole
momentum space. That is equivalent to assumming that $ \phi $ and $ \psi $
are two independent fields. Making such an assumption does not change the
effect
of the bath field greatly because in a realistic setting there exists
other environmental
fields which the system field interacts with (e.g., heat bath). Under
such an approximation the system field
is enhanced over the stochastic scheme in the high frequency sector,
but the  overall behavior of galaxy spectrums will not be affected
significantly, because it is
determined mainly by the low-frequency sector anyway.

Since the de Sitter space (3.1) is  conformally- flat, a changeover to
conformal time and conformally-related fields
$$
\tau=\int dt{1\over a(t)}
                                                        \eqno(3.13)
$$
$$
\tilde\psi(\vec x,\tau)
=a^{1-{n\over 2}}(\tau)\psi(\vec x)
                                                        \eqno(3.14)
$$
\noindent can simplify the calculations. Let us also define the conformal
mass by
$$
\tilde M^2=a^2 M^2
                                                        \eqno(3.15)
$$
\noindent It is clear that all the Feynman diagrams in (3.12) after the
conformal transformation are identical to those in (2.25)
%change eqno
which we have
calculated before. We find the following effective action (henceforth
$t$ will denote the conformal time $\tau$)

$$
\eqalign{
A[\phi,\phi']
& =S_r[\phi]
  -\int d^4x a^3(t)~{1\over 2}\delta m^2(t)\phi^2 \cr
& +\int d^4x a^3(t) \int d^4x' a^3(t')~{1\over 2}
   \lambda^2 \phi^2(x)V(x-x')\phi^2(x') \cr
& -S_r[\phi']
  +\int d^4x a^3(t){1\over 2}\delta m^2(t)\phi'^2\Bigr\} \cr
& -\int d^4x a^3(t) \int d^4x' a^3(t')~{1\over 2}
   \lambda^2 \phi'^2(x)V(x-x')\phi'^2(x') \cr
& -\int\limits_{t_0}^{t_f}dt\int d^3\vec x a^3(t)
   \int\limits_{t_0}^tdt'\int d^3\vec x' a^3(t')
   \lambda^2\Bigl[\phi^2(x)-\phi'^2(x)\Bigr] \cr
& ~~~~~ \times
   \eta(x-x')\Bigl[\phi^2(x')+\phi'^2(x')\Bigr] \cr
& +i\int\limits_{t_0}^{t_f}dt\int d^3\vec x a^3(t)
   \int\limits_{t_0}^tdt'\int d^3\vec x' a^3(t')
   \lambda^2\Bigl[\phi^2(x)-\phi'^2(x)\Bigr] \cr
& ~~~~~ \times
   \nu(x-x')\Bigl[\phi^2(x')-\phi'^2(x')\Bigr] \cr}
                                                        \eqno(3.16)
$$

\noindent where
$$
V(x-x')=\mu(x-x')-sgn(t-t')\eta(x-x')
                                                        \eqno(3.17)
$$

\noindent is the kernel of the non-local potential.
Here we have introduced the counter terms for the mass, the field-geometry
coupling constant, and the self-interaction coupling constant renormalization
of the $ \phi $ and $ \phi' $ fields respectively, and with them the
corresponding physical parameters.

As before, we see that the last two terms in (3.16) are the dissipation and
noise terms whose kernels are given by (2.28) and (2.29)
%CHANGE EQNO
respectively with
conformal time here replacing cosmic time in the Minkowsky space results.
The dissipation is of a nonlinear non-local type. The noise is coupled
to the system with an action in the form

$$
\int d^4x\sqrt{-g(x)}\Bigl\{ \xi (x)\phi^2(x) \Bigr\}
                                                        \eqno(3.18)
$$

\noindent  The stochastic force
(noise) $ \xi (x) $ has the following functional distribution

$$
P[\xi]
=N\times\exp \Biggl\{-{1\over 2}\int d^4x\int d^4x'~\xi(x)
 \Bigl[{\nu^{-1}(x-x')\over \lambda^2a^3(t)a^3(t')}\Bigr]~\xi(x')\Biggr\}
                                                        \eqno(3.19)
$$

\noindent One can show from this that
$$
\Biggl\{
\eqalign{
& <\xi(x)>_{\xi}=0 \cr
& <\xi(x)\xi(x')>_{\xi}=\nu(x-x') \cr}
                                                        \eqno(3.20)
$$

\noindent So this is a nonlinearly-coupled colored noise.
The fluctuation-dissipation relation for this field model
in de Sitter space is exactly the same as that in Minkowski space (2.37)
and (2.38).
%CHANGE EQNO OK
%Comparing the above results with Starobinsky's work, we can find many major
%differences. In their work, there was no dissipation at all, and the noise
%source is a linearly coupled white noise. However, if we take into account
%the mode-mode coupling of the inflation field, there is both noise and
%dissipation. The noise is not white and it is non-linearly coupled to the
%long wavelength modes. The dissipation is also nonlinear non-local
%%dissipation.
%More importantly, there exists a fluctuation-dissipation relation between
%the noise and dissipation. One also can show that in the classical limit,
%the quantum dissipation and noise reduce to their classical counterparts.

This sample calculation shows the origin and nature of noise from a
quantum field in a cosmological setting. We can now turn to the second
issue raised at the beginning, i. e., decoherence in the long wave-length
sector. To analyse this problem we need to know  the master equation,
at least the form of the diffusion terms in that equation.

The functional quantum master equation for this field-theoretical
model with general
nonlinear non-local dissipation and non-linearly coupled colored noise has
a complicated form in cosmic time ( denoted before as $ t$) .
However, in conformal time (in these equations also denoted as $t $),
it is similar to that in Minkowsky spacetime, which has been
derived in \refto{HPZ2}. We will not repeat that derivation here, but
just mention a simple case to end our discussion. This is the case
in cosmic time where one can get an explicit
form of the functional quantum master equation, i.e., by making a
local truncation in the effective action (3.16). Setting

$$
V(x-x')=v_0(t)\delta^4(x-x')
                                                        \eqno(3.21)
$$
$$
\eta(x-x')
={\partial\over\partial(t-t')}
\Bigl\{\gamma_0(t)\delta(x-x')\Bigr\}
                                                        \eqno(3.22)
$$
$$
\nu(x-x')=\nu_0(t)\delta(x-x')
                                                        \eqno(3.23)
$$
\noindent we get the effective action
$$
\eqalign{
& A[\phi,\phi'] \cr
& =\int\limits_0^tds\int d^3\vec x\Biggl\{
   {1\over 2}\dot\phi^2
  -{1\over 2}{1\over a^2(t)}\big[\nabla\phi\big]^2
  -{1\over 2}\Bigl[m^2_r+{1+\xi_r\over 6}R(x)\Bigr]\phi^2
  -{1\over 4!}\lambda_r\phi^4 \cr
& -{1\over 2}\dot\phi'^2
  +{1\over 2}{1\over a^2(t)}\big[\nabla\phi'\big]^2
  -{1\over 2}\Bigl[m^2_r+{1+\xi_r\over 6}R(x)\Bigr]\phi'^2
  +{1\over 4!}\lambda_r\phi'^4 \cr
& -{1\over 2}\delta m^2(t)\phi^2
  +{1\over 2}\delta m^2(t)\phi'^2
  +{1\over 2}\lambda^2 v(t)\phi^4
  -{1\over 2}\lambda^2 v(t)\phi'^4 \cr
& -2\lambda^2a^3(t)\gamma_0(t)(\phi^2-\phi'^2)
   (\phi\dot\phi-\phi'\dot\phi')
  -3\lambda^2a^2(t)\dot a(t)\gamma_0(t)(\phi^4-\phi'^4) \cr
&
  +i\lambda^2\nu_0(t)(\phi^2-\phi'^2)^2 \Biggr\}\cr }
                                                        \eqno(3.24)
$$

\noindent From this we can derive
the functional quantum master equation in the local truncation approximation
\refto{Zhang}:
$$
i{\partial\over\partial t}~\rho_r[\phi,\phi',t]
=\hat H_{\rho}[\phi,\phi',t]~\rho_r[\phi,\phi',t]
                                                        \eqno(3.25)
$$
\noindent where
$$
\eqalign{
H_{\rho}[\phi,\phi',t]
& =\int d^3\vec x a^3(t)\Biggl\{
   \hat h_r(\phi)-\hat h_r(\phi') \cr
&  +3\lambda^2a^2(t)\dot a(t)\gamma_0(t)
   \Bigl[\phi^4(\vec x)-\phi'^4(\vec x)\Bigr] \cr
& +2\lambda^2\gamma_0(t)
   \Bigl[\phi^2(\vec x)-\phi'^2(\vec x)\Bigr]
   \Bigl[\phi(\vec x){\delta\over\delta\phi(\vec x)}
   -\phi'(\vec x){\delta\over\delta\phi(\vec x)}\Bigr] \cr
& -i\lambda^2\nu_0(t)\Bigl[\phi^2(\vec x)
   -\phi'^2(\vec x)\Bigr]\Biggr\} \cr}
                                                        \eqno(3.26)
$$
\noindent and
$$
\eqalign{
\hat h_r(\phi)=
& -{1\over 2}{1\over a^6(t)}{\delta^2\over\delta\phi^2(\vec x)}
  +{1\over 2}a(t)\big[\nabla\phi(\vec x)\big]^2
  +{1\over 2}\Bigl[m^2_r+{1+\xi_r\over 6}R(t)\Bigr]\phi^2(\vec x) \cr
& +{1\over 4!}\lambda_r\phi^4(\vec x)
  +\delta m^2(t)\phi^2(\vec x)
  -{1\over 2}\lambda^2v(t)\phi^4(\vec x) \cr }
                                                         \eqno(3.27)
$$

This functional quantum master equation and its associated
Langevin equation or Fokker-Planck-Wigner equation can be used to analyze
the dynamics of the system field (long wavelength modes in the stochastic
inflation scheme) for studying the  decoherence and structure formation
processes in the early universe. We have only begun this investigation
and details will be made available in a later publication.% \refto{HP}.
Here,
as a preliminary result, we can get some qualitative information on how the
system decoheres by analyzing the behavior of the diffusion term in the
master equation. Diffusive effects are
generated by the last term in the effective
action (3.16) that produces the following contribution on
the right hand side of the master equation for
$\rho[\phi, \phi']$:
$$
\dot\rho[\phi, \phi', t]\propto-\bigl(\phi^2-\phi'^2\bigr)*D(t)*
\bigl(\phi^2-\phi'^2\bigr) \times\rho[\phi, \phi',t]       \eqno(3.28)
$$
\noindent Here the symbol $*$ denotes the convolution product and
$\phi$ represents a configuration of the scalar
field in a surface of constant conformal time.
The diffusion "coefficient'' $D$ is
therefore a nonlocal kernel that can be written in terms of its spatial
Fourier transform as
$$
D(\vec x, \vec y, t)=\int {{d\vec k}\over {(2\pi)^3}} \nu_{\vec k}(t)~
\exp\bigl(-i\vec k \dot (\vec x - \vec y)\bigr)                  \eqno(3.29)
$$
It is not easy to analyze the effect produced by a term like the one appearing
in (3.28). However, we can use the following argument (see \Ref{PazTFT})
to  qualitatively
investigate if the diffusive effects are stronger for long wavelength modes
than they are for short ones.  Note that the coefficient
in (3.28) can be written in terms of the product of the Fourier
transform (3.29) and that of the field $Q(x)=\phi^2$:
$$
\bigl(\phi^2-\phi'^2\bigr)*D(t)*\bigl(\phi^2-\phi'^2\bigr)=
\int d\vec k (Q-Q')_{\vec k} D_{\vec k} (Q-Q')_{\vec k}            \eqno(3.30)
$$

Let us now examine the dependence on
$ k=|\vec k |$ of the function $D_{\bf k}$ entering
in (3.29). Using our previous results, it is not hard to prove that this
function can be written in terms of the physical wave vector
${\bf p}={\bf k}/a$ as
$$
D_{\bf k}(t)={{a^4}\over{4\pi}}\lambda^2\Bigl(1-{{H}\over{{\bf p}}}
f({{\bf p}\over H}) + g({{\bf p}\over H})\Bigr)                  \eqno(3.31)
$$
where
$$
\eqalignno{
f(x)& = {1\over{2\pi}} \int_0^{2x} dx
[-\sin x ~ Ci(x) + \cos x Si (x)] &                  (3.32.a)\cr
g(x)& = {1\over{2\pi}} \int_0^{2x} dx
[\cos x Ci (x) + \sin x Si (x)]   &                  (3.32b)\cr}
$$
\noindent and $Si(x),~Ci(x)$ are the usual integral trigonometric functions.
In Figure 1 we have plotted $D_{\bf k}(t)$ for a fixed value of
the conformal time as a function of ${{\bf p}\over H}$, i.e.,
the ratio between the horizon size and the physical wavelength.
The function has a strong peak in the infrared region of the spectrum
suggesting that
diffusion effects (decoherence is one of them) are indeed more pronounced
for long wavelength modes and  weaker
for wavelengths shorter than the horizon size.

%\vskip 2cm
\vfill
\eject

\noindent{\bf 4. Discussions}

We have outlined the first part of a program to describe
structure formation from primordial quantum fluctuations
via stochastic dynamics. To end,
we briefly summarize our findings and discuss
the feasibility of our mechanism and its implications.

{\it A. What is new?}

What we have accomplished here are:
1) supply a quantum field-theoretical definition and derivation of
noise; 2) relate different types of noise to different couplings of
the system and environment; and 3) derive an equation of motion
--the master equation for the reduced density matrix or the
Fokker-Planck equation for the associated Wigner functional.
In this process
4) we showed from first principles how one can derive the equations of
stochastic dynamics
from interacting quantum field theory, both in flat and curved spacetimes;
and 5) we proposed a new scheme of noise generation based on nonlinear
coupling which is different from the
Starobinsky mechanism (which assumes a free field with a moving partition).
Since an interacting field (e.g., a $ \lambda\phi^4 $ or
Coleman-Weinberg potential) is what is usually used for generating
inflation anyway our mechanism is rather natural.

{\it B. How realistic are the conditions?}

The noise in our scheme arises from the system field ( the inflaton)
coupling nonlinearly to an environment field. What in a realistic situtation
could play the role of the environment field?
One can assume as in the stochastic inflation scheme that the system field
consists of the low frequency modes and the environment field that of
the high frequency modes of one single inflaton field.
The model we have studied,  which has two
separate self-interacting  scalar fields coupled biquadratically
each assuming a full spectrum of modes, can be viewed as an approximation
to this scheme. The environment field can also be referring to other
fields present besides the inflaton field. Only the quantum fluctuations
of such fields need be present in our scheme to generate the noise which
seeds the galaxies. Even if one assumes nothing,
there is always the gravitational
field itself which the inflaton field is coupled to, and the vacuum
gravitational fluctuations can equally seed the structures in our
universe \refto{Gri, MFB}.
(Note that in such cases the coupling is of a derivative form
rather than the polynomial form in this example. Noise arising from a
derivative type of coupling has been studied in connection with the
issue of gravitational entropy in minisuperspace quantum cosmology
\refto{HuWaseda}.)

{\it C. Physical consequences}

Noises arising from nonlinear couplings are under general circumstances
colored.
They generate fluctuations which could give rise to non-Gaussian
galaxy distributions (NGD). There are, of course, simpler ways to generate NGD.
A changing Hubble rate $H= \dot a/a$ as in a `slow-roll' transition, or an
exponential potential $V(\phi)$ \refto{LucMat}
will do. However, such mechanism only generates NGD at very long wavelengths,
much longer than the horizon size to be relevant to the observable spectrum.

As for the present scheme, since the value of $\lambda$ is restricted to be
very small ($< 10^{-12}$) in the standard GUT inflationary models (so that the
magnitude of the density contrast is compatible with the observed value
$\delta \rho / \rho \approx 10 ^{-4}$ when the fluctuation mode enters the
horizon), the constituency of the colored portion of the noise is accordingly
small. Whether a nonlinear coupling will generate excessive inhomogeneities is
an open question. It is still too premature for us to speculate on the
general behavior. Details of galaxy formation analysis from the stochastic
equations of motion derived here with different types of colored noise and
realistic physical parameters will be reported in a later publication.
%\refto{HP}.

\vfill
\eject

\references
%\input referencefile
%Bela.bib: Bibliography of Belgium talk (1992)  Bela.tex

\refis{Lifshitz}  E. M. Lifshitz and I. Kalatnikov, Adv. Phys. 12, 185
(1963).

\refis{Bardeen}  J. Bardeen, Phys. Rev. {\bf D 22}, 1882 (1980).
G. Ellis and M. Bruni, Phys. Rev. D40, 1804 (1989)

\refis{Peebles}  P. J. E. Peebles, {\it Large Scale Structure of the
Universe} (Princeton University Press, Princeton, 1980)

\refis{ZelNov}  Ya. B. Zel'dovich and I. D. Novikov, {\it Relativistic
Cosmology} Vol. 2, (University of Chicago Press, Chicago, 1985)

\refis{Guth}  A. H. Guth, Phys. Rev. D 23, 347 (1981).

\refis{AlbSte}  A. Albrecht and P. J. Steinhardt, Phys. Rev. Lett. 48,
1220 (1982).

\refis{LindeInf}  A. D. Linde, Phys. Lett. 114B, 431 (1982).

\refis{GalForInf}  A. Guth and S. Y. Pi, Phys. Rev. Lett. {\bf 49}, 1110
(1982). A. A. Starobinsky, Phys. Lett. {\bf 117B}, 175 (1982). S. W.
Hawking, Phys. Lett. {\bf 115B}, 295 (1982). J. M. Bardeen, P. J. Steinhardt
and M. S. Turner, Phys. Rev. {\bf D28}, 629 (1983); R. Brandenberger, R.
Kahn and W. Press, Phys. Rev. {\bf D 28}, 1809 (1983).

\refis{LindeChaos}  A. Linde, Phys. Lett. 162B, 281 (1985).

\refis{LucMat}  F. Lucchin and S. Matarrese, Phys. Rev. D32, 1316 (1985).

\refis{GuthPiInf}  A. H. Guth and S.- Y. Pi, Phys. Rev. D32, 1899 (1985).

\refis{Staro86} A. A. Starobinsky, in {\it Field Theory, Quantum Gravity
and Strings}, ed. H. J. de Vega and N. Sanchez (Springer, Berlin 1986).

\refis{BarBub}
J. M. Bardeen and G. J. Bublik, Class. Quan. Grav. {\bf 4}, 473 (1987).

\refis{Rey} S. J. Rey, Nucl. Phys. B284, 706 (1987).

\refis{Graziani}  F. Graziani, Phys. Rev. D38, 1122, 1131, 1802 (1988);
  D39, 3630 (1989).

\refis{envdec}
W. H. Zurek, Phys. Rev. {\bf D24}, 1516 (1981); {\bf D26}, 1862 (1982);
in {\it Frontiers of Nonequilibrium Statistical Physics}, ed. G. T.
Moore and M. O. Scully (Plenum, N. Y., 1986); Physics Today {\bf 44}, 36
(1991);
E. Joos and H. D. Zeh, Z. Phys. {\bf B59}, 223 (1985); A. O. Caldeira and A. J.
Leggett, Phys. Rev. {\bf A31}, 1059 (1985);
W. G. Unruh and W. H. Zurek, Phys. Rev. {\bf D40}, 1071 (1989).
B. L. Hu, J. P. Paz and Y. Zhang, Phys. Rev. {\bf D45}, 2843 (1992);
{\bf D47}, 1576 (1993);
J. P. Paz, S. Habib and W. H. Zurek, Phys. Rev. {\bf D47}, 488 (1993).
W. H. Zurek, J. P. Paz and S. Habib, Phys. Rev. Lett. {\bf 70}, 1187 (1993);
W. H. Zurek, Prog. Theor. Phys. {\bf 89}, 281 (1993).

\refis{conhis}
R. B. Griffiths, J. Stat. Phys. {\bf 36}, 219 (1984); R. Omn\'es, J. Stat Phys.
{\bf 53}, 893, 933, 957 (1988); Ann. Phys. (N. Y.) {\bf 201}, 354 (1990); Rev.
Mod. Phys. {\bf 64}, 339 (1992); {\it The Interpretation of Quantum Mechanics},
(Princeton University Press, Princeton (1994)). J. B. Hartle, ``Quantum
Mechanics of Closed Systems'' in {\it Directions in General Relativity} Vol.
1, eds B. L. Hu, M. P. Ryan and C. V. Vishveswara (Cambridge Univ.,
Cambridge, 1993); M. Gell-Mann and J. B. Hartle, in {\it Complexity, Entropy
and the Physics of Information}, ed. by W. H. Zurek (Addison-Wesley,
Reading, 1990); J. B. Hartle and M. Gell- Mann, Phys. Rev. {\bf D47}, 3345
(1993).  J. P. Paz and S. Sinha, Phys. Rev. {\bf D44}, 1038 (1991).
H. F. Dowker and J. J. Halliwell, Phys. Rev. {\bf D46}, 1580 (1992).
T. Brun, Phys. Rev. {\bf D47}, 3383 (1993).
J. Twamley, Phys. Rev. {\bf D48}, 5730 (1993).
J. P. Paz and W. H. Zurek, Phys. Rev. {\bf D48}, 2728 (1993).

\refis{decQC}  C. Kiefer, Clas. Quant.
Grav. {\bf 4}, 1369 (1987); J. J. Halliwell, Phys. Rev. {\bf D39}, 2912
(1989); T. Padmanabhan, {\it ibid.} 2924 (1989). B. L. Hu ``Quantum and
Statistical Effects in Superspace Cosmology'' in {\it Quantum Mechanics in
Curved Spacetime}, ed. J. Audretsch and V. de Sabbata (Plenum, London 1990)
E. Calzetta, Class. Quan. Grav. {\bf 6}, L227 (1989); Phys. Rev. {\bf D 43},
2498 (1991); J. P. Paz and S. Sinha, Phys. Rev. {\bf D44}, 1038 (1991);
{\it ibid} {\bf D45}, 2823 (1992); B. L. Hu, J. P. Paz and S. Sinha,
''Minisuperspace as a Quantum Open System'' in {\it Directions in General
Relativity} Vol. 1, (Misner Festschrift) eds B. L. Hu , M. P. Ryan and C. V.
Vishveswara (Cambridge Univ., Cambridge, 1993)

\refis{BraLafMij}
R. Brandenberger, R. Laflamme and M. Mijic, Physica Scripta T36, 265 (1991);
R. Laflamme and A. Matacz, Int. J. Mod. Phys. D2, 171 (1993).

\refis{HuZhaUnc}  B. L. Hu and Yuhong Zhang, ``Quantum and Thermal
Fluctuations, Uncertainty Principle, Decoherence and Classicality'' in Proc.
Third International Workshop on Quantum Nonintegrability, Drexel University,
Philadelphia, May 1992, eds J. M. Yuan, D. H. Feng nd G. M. Zaslavsky
(Gordon and Breach, Langhorne, 1993). B. L. Hu and Yuhong Zhang, Mod. Phys.
Lett. A8, 3575 (1993); Univ. Maryland preprint 93-162 (1993), gr-qc/93012034

\refis{nonGaussian}
D. Salopek, J. D. Bond and J. M. Bardeen,  Phys. Rev. D40, 1753 (1989);
A. Ortolan, F. Lucchin and S. Mataresse,  Phys. Rev. D38, 465 (1988);
S. Mataresse and A. Ortolan,  Phys. Rev. D40, 290 (1989);
H. M. Hodges, Phys. Rev. D39, 3568 (1989);
I. Yi, E. T. Vishniac and S. Mineshige,  Phys. Rev. D43, 362 (1991).

\refis{HuTsukuba}
B. L. Hu, ``Statistical Mechanics and Quantum Cosmology'',
in {\it Proc. Second International Workshop on Thermal Fields
and Their Applications}, eds. H. Ezawa et al (North-Holland, Amsterdam,
1991).

\refis{HPZ1}
B. L. Hu, J. P. Paz and Y. Zhang, Phys. Rev. {\bf D45}, 2843 (1992)

\refis{HPZ2}
B. L. Hu, J. P. Paz and Y. Zhang, Phys. Rev. {\bf D47}, 1576 (1993)

\refis{ctp}  J. Schwinger, J. Math. Phys. {\bf 2} (1961) 407; P. M. Bakshi
and K. T. Mahanthappa, J. Math. Phys. 4, 1 (1963), 4, 12 (1963). L. V.
Keldysh, Zh. Eksp. Teor. Fiz. {\bf 47 }, 1515 (1964) [Engl. trans. Sov.
Phys. JEPT {\bf 20}, 1018 (1965)]; G. Zhou, Z. Su, B. Hao and L. Yu, Phys.
Rep. {\bf 118}, 1 (1985); Z. Su, L. Y. Chen, X. Yu and K. Chou, Phys. Rev.
{\bf B37}, 9810 (1988); B. S. DeWitt, in {\it Quantum Concepts in Space and
Time} ed. R. Penrose and C. J. Isham (Claredon Press, Oxford, 1986); R. D.
Jordan, Phys. Rev. {\bf D33 }, 44 (1986). E. Calzetta and B. L. Hu,
{\it Phys. Rev.} {\bf D35}, 495 (1987).

\refis{CH87}
E. Calzetta and B. L. Hu, Phys. Rev. {\bf D35}, 495 (1987).

\refis{cgea}  B. L. Hu and Y. Zhang, ``Coarse-Graining, Scaling, and
Inflation'' Univ. Maryland Preprint 90-186 (1990); B. L. Hu, in {\it
Relativity and Gravitation: Classical and Quantum} Proceedings of SILARG VII,
Cocoyoc, Mexico, Dec. 1990. eds. J. C. D' Olivo {\it et al}
(World Scientific, Singapore, 1991).

\refis {SinHu}
Sukanya Sinha and B. L. Hu, Phys. Rev. D44, 1028 (1991)

\refis{CorBru}  J. M. Cornwall and R. Bruinsma, Phys. Rev. D38, 3146 (1988);
M. Sagakami, Prog. Theor. Phys. 79, 443 (1988);

\refis{Zhang}
Yuhong Zhang, Ph. D. Thesis, University of Maryland (1990)

\refis{PazTFT}
J. P. Paz, in {\it Proc. Second International Workshop on Thermal Fields
and Their Applications}, eds. H. Ezawa et al (North-Holland, Amsterdam, 1991).

\refis{Gri}
L. Grishchuk and Y. V. Sidorov, Phys. Rev. D42, 3414 (1990)

\refis{MFB}
V. Mukhanov, H. Feldman and R. Brandenberger, Phys. Rep. {\bf 215}, 203 (1992).

\refis{HuWaseda}
B. L. Hu, ``Quantum Statistical Processes in the
Early Universe'' in {\it Quantum Physics and the Universe}, Proc. Waseda
Conference, Aug. 1992 eds. M. Namiki et al (Pergamon Press, Tokyo, 1993).
Vistas in Astronomy 37, 391 (1993). gr-qc/9302031

\endreferences
\end